\documentclass[elsart12,epsf,eqsecnum,graphics,cite]{revtex4}

\usepackage{epsfig,epsf}
\usepackage{graphicx, subfigure}
\usepackage{graphicx}
\usepackage{grffile}
\newcommand{\be}{\begin{equation}}
\newcommand{\ee}{\end{equation}}

\begin{document}


\voffset1.5cm

\title{Some Comments on Ghosts and Unitarity: The Pais-Uhlenbeck Oscillator Revisited.}
\author{ Ibrahim Burak Ilhan$^{1}$ and Alex Kovner$^{1}$}

\affiliation{
$^1$ Physics Department, University of Connecticut, 2152 Hillside
Road, Storrs, CT 06269-3046, USA}

\begin{abstract}
 We give a simple discussion of  ghosts, unitarity violation, negative norm states and quantum vs classical behavior in the simplest model with four derivative action - the Pais-Uhlenbeck oscillator. We also point out that the normalizable ``vacuum state'' (in the sense defined below) of this model can be understood as spontaneous breaking of the emergent conformal symmetry. We provide an example of an interacting system that couples the ``particle" and ``ghost" degrees of freedom and nevertheless remains unitary on both classical and quantum level.
\end{abstract}
\maketitle
\section{Introduction}
The physics of systems with ghosts has recently attracted renewed attention\cite{ghosts}. 
The most interest in these systems is in connection with the theories of gravity. In particular the so called $W^2$ gravity, the theory with local conformal symmetry is known to have ghost modes\cite{weyl}. This is usually considered to be a hindrance for a physical theory. Indeed an absence of a stable vacuum (lowest energy) state is disconcerting and is likely to lead to an instability, whereby the evolution extracts energy from the negative energy modes and pumps it into the positive energy modes producing a runaway instability. 

On the other hand, conformal gravity possesses much improved UV properties compared to the Einstein gravity, which render it renormalizable by power counting. The gravitational force in the theory of conformal gravity at large distances differs from the Newtonian gravitational force and this modification is capable of explaining data on galactic rotational curves with one fitting parameter, without introducing of the dark matter hypothesis\cite{mannheim}. It has also been suggested that conformal gravity may be able to solve the cosmological constant problem\cite{philip}.

These attractive features of the theory prompted attempts to solve the ghost problem. One approach attempts to separate the ghost modes from the positive norm gravitons and ban their propagation ``by hand"\cite{ghosts}. Another attempt is to quantize the theory using a nonstandard definition of a quantum mechanical norm\cite{philip_carl},\cite{pt} following a more general program of quantizing PT invariant but non hermitian Hamiltonians\cite{pt1}. In a free limit this is essentially equivalent to treating the ghost modes as purely imaginary, which  flips the sign of the ghost part of the Hamiltonian. It is not as yet clear whether any of these approaches can work in the full interacting theory.

On the other hand, the instability in question may not be necessarily a fatal flaw. This is especially so in a theory of gravity, which governs the evolution of the universe and thus never actually relaxes to its ground state. Thus the nonexistence of a ground state in gravity may be just a way of life. In particular it has been suggested that a negative pressure due to ghosts may be a cause of the cosmological acceleration\cite{kaplan}. It has been argued that the time scale in which instability develops is way too short in theories which contain ghosts in the matter sector\cite{jeon}. We are unaware however of a similar analysis of gravitational ghosts themselves, that is the ghost partners of the gravitons that arise in conformal gravity: the rapid decay of the vacuum discussed in\cite{jeon} may be preventable if the ghost coupling to gravitons is nonlocal\cite{vilenkin}. It is not obvious therefore that the last word on viability of theories with ghosts has been uttered yet.

The purpose of this note  is rather modest and pedagogical. 
The potential interest notwithstanding, theories with ghosts are still considered somewhat esoteric and are not frequently discussed in particle physics literature. We aim to discuss pedagogically the simplest example of a theory with ghosts - the Pais-Uhlenbeck oscillator\cite{pais}. Our goal is to explicitly demonstrate in this simple framework the meaning of some rather paradoxical notions that are sometimes used in the context of theories with ghosts, like negative norm states and violation of unitarity in theories with ostensibly perfectly hermitian Hamiltonian.
We also demonstrate explicitly by solving the time dependent evolution in this theory how the soft UV behavior arises in dynamical context.

 We stress that the Pais-Uhlenbeck oscillator is in fact a unitary theory even though it possesses a ghost mode, and also give an example of a theory of interacting ``particle" and ``ghost" modes which is nevertheless unitary on the quantum level. All the above statements apply to quantum mechanical systems with the standard Dirac norm, as we do not recourse to a non standard quantization approach {\it a la} \cite{pt}.
 
The Pais - Uhlenbeck oscillator was the subject of several papers in recent years, and its solution is well known\cite{philip_carl,philip_aharon,smilga}. Nevertheless we feel that our simple and straightforward approach to the problem is illuminating and is worth recording.

\section{The Pais-Uhlenbeck oscillator}
The Pais-Uhlenbeck system is the theory of a single degree of freedom which satisfies a fourth order equation of motion. It is defined by the Lagrangian
\begin{equation}\label{lagr}
L=(\frac{d^2}{dt^2}z+\omega_1^2z)(\frac{d^2}{dt^2}z+\omega^2_2z)
\end{equation}
For definiteness we assume $\omega_1>\omega_2$.
Our aim is to study the Hamiltonian dynamics with the view of quantum mechanical system, since a discussion of evolution of a wave function is most convenient in the Hamiltonian formalism. 
Although there exist a general formalism for calculating a Hamiltonian of four derivative systems, developed by Ostragradsky\cite{ostrogradsky}, we find it more straightforward in the context of this particular model to introduce a pair of  variables
\begin{equation}\label{variables}
X=\frac{d^2}{dt^2}z+\omega^2_1z; \ \ \ \ \ Y=\frac{d^2}{dt^2}z+\omega^2_2z
\end{equation}
and consider them as independent coordinates. The rationale of this choice is, that the fourth order equation for the variable $z$
\begin{equation}
\left(\frac{d^2}{dt^2}+\omega_1^2\right)\left(\frac{d^2}{dt^2}+\omega_2^2\right)z=0
\end{equation}
can be written as a pair of second order equations for $X$ and $Y$
\begin{equation}
\frac{d^2}{dt^2}X+\omega^2_2X=0; \ \ \ \ \ \frac{d^2}{dt^2}Y+\omega^2_1Y=0
\end{equation}
To find the Hamiltonian we introduce the Lagrange multipliers for the constraints eq.(\ref{variables})
\begin{equation}
L=XY+\alpha(\frac{d^2}{dt^2}z+\omega_1z-X)+\beta(\frac{d^2}{dt^2}z+\omega_2z-Y)
\end{equation}
Canonical momenta are calculated in the standard fashion $p_i=\partial L/\partial \dot x_i$. This definition leads to the following constraints 
\begin{equation}\label{primary}
P_X=P_Y=0; \ \ \ p_\alpha=p_\beta=-\dot z ; \ \ \ \ \ p_z=-\dot\alpha-\dot\beta
\end{equation}
The Hamiltonian, calculated in the standard way as the Legendre transform of the Lagrangian is
\begin{equation}
H=-\frac{1}{2}(p_\alpha+p_\beta)p_z-XY-\alpha(\omega_1^2z-X)-\beta(\omega_2^2z-Y)
\end{equation}
Commuting (calculating the Poisson bracket of) $H$ with the primary constraints, eq.(\ref{primary}) we obtain secondary constraints
\begin{equation}
[H, P_X]=\alpha-Y=0; \ \ [H, P_Y]=\beta-X=0; \ \ \ [H, p_\alpha-p_\beta]=(\omega_1^2-\omega_2^2)z-(X-Y)=0
\end{equation}
These can be used  to express $\alpha,\ \beta$ and $z$ in terms of $X$ and $Y$,
\begin{equation}
\alpha=Y; \ \ \beta=X; \ \ \ z=\frac{X-Y}{\omega^2_1-\omega^2_2}
\end{equation}
The Dirac procedure for constraint systems requires that we use the Dirac brackets instead of the Poisson brackets to derive equations of motion. The net result of switching to the Dirac brackets is clear without a detailed calculation. The new ``commutation relations" are such that the dynamical variables ``commute" with all the constraints. Also, the modification is present only for those variables whose Poisson bracket with the original constraints does not vanish. Without any calculation the result in the present case is obvious
\begin{equation}
p_\alpha=\pi_Y; \ \ p_\beta=\pi_X; \ \ p_z=\frac{1}{2}(\omega^2_1-\omega^2_2)(\pi_X-\pi_Y)
\end{equation}
with the Dirac brackets 
\begin{equation}
[\pi_i,X_j]_D=-\delta_{ij}
\end{equation}
The Hamiltonian then becomes
\begin{equation}
H=\frac{1}{2}\Omega\Delta[\pi_Y^2-\pi_X^2]+\frac{\omega_1^2}{2\Omega\Delta}Y^2-\frac{\omega_2^2}{2\Omega\Delta}X^2
\end{equation}
where we have defined
$\Omega=\frac{\omega_1+\omega_2}{2}; \ \ \ \Delta=\omega_1-\omega_2$.
Finally rescaling the variables
$\pi_x=(\Omega\Delta)^{1/2}\pi_X;\ \ \ \  x=(\Omega\Delta)^{-1/2}X$, and similarly for $y$ we obtain
\begin{equation}\label{hamxy}
H=\frac{1}{2}\pi_y^2+\frac{1}{2}\omega_1^2y^2-\frac{1}{2}\pi_x^2-\frac{1}{2}\omega_2^2x^2
\end{equation}
In terms of the new variables, the original coordinate $z$ is expressed as
\begin{equation}
z=\frac{1}{\sqrt {4\Omega\Delta}}(x-y)
\end{equation}

This is a very simple Hamiltonian. It is not bounded either from below nor from above, but nevertheless it generates a perfectly acceptable evolution. The two degrees of freedom $x$ and $y$ are decoupled, and classically each one simply satisfies a harmonic oscillator equation of motion. Classically  there are no runaway solutions for these equations of motion for an arbitrary initial condition. Quantum mechanically the system also possesses finite positive norm states which evolve unitarily with time.

Nevertheless the common jargon  is that this quantum theory has negative norm states. In the next subsection we will clarify what this statement technically means, and stress that it is not a hindrance for a peaceful existence of a unitary evolution in this model.  

\subsection{The Negative Face of a Divergent Integral}
The Hamiltonian of the Pais-Uhlenbeck system is not bounded from below. This is unusual and somewhat disturbing, since we normally expect that any open system will interact with some external degrees of freedom and generally relax to its ground state by loosing any excess energy to those degrees of freedom. However if the system is closed, no such loss of energy is possible and unboundedness of energy from below does not have to be a problem. In particular, in the present case the two harmonic oscillators do not interact with each other, no energy transfer from one to another occurs and the evolution is perfectly unitary, provided at the initial moment in time we start with a state which is localized at finite values of $x$ and $y$.

If one does insist, however to push the system to the lowest energy state, the evolution becomes non-unitary. This is simply due to the fact that this state is localized at infinite values of $x$, is non-normalizable and thus the probability ``leaks" through the spatial boundary. 

To see this explicitly, let us define creation and annihilation operators in the standard way
\begin{equation}
a=\sqrt{\frac{\omega}{2}}x+i\sqrt{\frac{1}{2\omega}}\pi_x
\end{equation}
The Fock vacuum of $a$ is the normalized Gaussian state
\begin{equation}
a|0\rangle=0; \ \ \ \ \ \langle x\vert 0\rangle=Ne^{-\frac{\omega}{2}x^2}
\end{equation}
This is the state with highest energy in the $x$-sector. 

One can also formally define a state which corresponds to lowest energy eigenvalue, as the vacuum of $a^\dagger$\cite{philip_aharon}
\begin{equation}
a^\dagger|\Phi\rangle=0; \ \ \ \ \ \langle x\vert\Phi\rangle=N_-e^{\frac{\omega}{2}x^2}
\end{equation}
This state is non-normalizable and not physical, since a particle in this state is localized exclusively at infinity. The probability to find the particle at finite value of coordinate vanishes, since in the infinite volume limit the normalization constant $N_-$ vanishes faster than exponentially.

 Nevertheless in a certain formal way it corresponds to the lowest energy state. To see this, write the Hamiltonian for $x$ mode in the standard form
\begin{equation}
H_x=-\omega aa^\dagger + E_0
\end{equation}
Consider a tower of states above $|\Phi\rangle$ generated by the action of operator $a$.
\begin{equation}\label{tower}
H|\Phi\rangle=E_0|\Phi\rangle; \ \ \ H|1\rangle\equiv Ha|\Phi\rangle=E_0a|\Phi\rangle-\omega a a^\dagger a|\Phi\rangle=(E_0+\omega)|1\rangle; \ \ ...
\end{equation}
Thus applying operator $a$ increases the energy of the state by $\omega$, and the spectrum seems to be bounded from below. 
Another formal argument suggests that at least some of these states have negative norm. Let us calculate the norm of the ``one particle state"
\begin{equation}
\langle1|1\rangle=\langle\Phi|a^\dagger a|\Phi\rangle=\langle\Phi|aa^\dagger-1|\Phi\rangle=-\langle \Phi|\Phi\rangle
\end{equation}
Taken literally, this argument suggests that either the ``one particle" state or the ``vacuum" state has a negative norm. 
This is the origin of the usual statement that the theory has negative norm states.

In fact, of course the norm of both of these states is positive once we regulate the system by putting it into a finite volume. The ``vacuum" state is just a Gaussian which grows at large values of $x$. Its norm is positive in finite volume, and diverges (while remaining positive) as the infrared cutoff is removed. The one particle wave function can be found explicitly
\begin{equation}\label{negnorm}
a|\Phi\rangle=\left(\sqrt{\frac{\omega}{2}}x+\sqrt{\frac{1}{2\omega}}\frac{d}{dx}\right)e^{\frac{\omega}{2}x^2}=\sqrt{2\omega}xe^{\frac{\omega}{2}x^2}
\end{equation}
The norm of this state obviously is also positive, and  is even more divergent than that of the vacuum in large volume. None of the norms is negative. The flaw in the formal eq.(\ref{negnorm}) is of course precisely the fact that the states in question are not normalizable. To interpret the expectation value of $a^\dagger a$ as the norm of a one particle state, one needs to act with $a^\dagger$ on the bra, which amounts to integration by parts of the derivative in $a^\dagger$. The integration by parts however is not allowed, since the wave function grows at infinity. In particular
\begin{equation}
\langle 1|1\rangle\ne |a|\Phi\rangle|^2
\end{equation} 
as one can easily verify by an explicit calculation. In fact the difference between the two sides of the inequality is infinite. Thus ``negative norm" is merely a jargon which refers to the fact that neither the norm nor matrix element of any reasonable operator like $x^n$ or $p^n$ is defined in the states of the form eq.(\ref{tower}) due to strong infrared divergence.

Sometimes the procedure described above is referred to as a ``quantization scheme", in the sense that the states of the tower eq.(\ref{tower}) do not belong to the Hilbert space of normalizable states. The unitarity in this quantization scheme is broken exactly for the reason explained above. All the wave functions with finite number of ``excitations" above the ``vacuum" $\vert\Phi\rangle$ live on the edge of space. Once an infrared regulator (which makes the norm finite) is removed the wave functions vanish everywhere in the bulk. Such states run great risk of disappearing through the boundary under time evolution.  

On the other hand it is clear, that states which are created by the action of $a^\dagger$ on  $\vert 0\rangle$ are normalizable and their evolution is perfectly unitary. One is normally interested in the situation when a particle can be detected in the bulk with finite probability. This physical condition makes the non-normalizable states physically irrelevant and devoid of interest.

\section{The degenerate case $\Delta=0$}
A special case of the Pais-Uhlenbeck system is  when the two oscillators have the same frequency, $\Delta=0$. In terms of analogy with the $W^2$ gravity, this case is the most interesting. In this section we discuss some interesting features of the equal frequency limit.
\subsection{The Fate of the Normalized Wave Functions}
The limit $\Delta=0$ of the previous expressions is a little tricky, since the transformation between the original variable $z$ and $x,y$ becomes singular. It is therefore not straightforward to take the limit directly on the level of the Hamiltonian. One cannot simply drop the terms in the Hamiltonian which naively vanish in the limit $\Delta\rightarrow 0$, since the operators that multiply $\Delta$ may have divergent matrix elements. To illustrate this, let us first rewrite the Hamiltonian in terms of variables $X$ and $z$, avoiding any singular redefinition of variables (here the variable $x$ is defined as originally: $X=\frac{d^2}{dt^2}z+\omega^2 z$).

\begin{equation}\label{hamxz}
H=-\frac{1}{2}\pi_X\pi_z+X^2-2(\Omega+\frac{1}{2}\Delta)^2zX+2\Delta\Omega(\Omega+\frac{1}{2}\Delta)^2z^2-\frac{1}{2}\Delta\Omega\pi_X^2
\end{equation}

Suppose we naively  drop the last two terms in eq.(\ref{hamxz}), which formally vanish in the limit $\Delta\rightarrow 0$. 
\begin{equation}\label{hamxz1}
H_0=-\frac{1}{2}\pi_X\pi_z+X^2-2(\Omega+\frac{1}{2}\Delta)^2zX
\end{equation}
Let us now look for Gaussian eigenstates of the resulting Hamiltonian. Recall that at nonzero $\Delta$ we had four Gaussian eigenstates 
\begin{equation}\label{states}
\exp \pm\left\{\frac{\omega_2}{2\Omega\Delta}X^2\pm \frac{\omega_1}{2\Omega\Delta}Y^2\right\}=\exp\pm\left\{\frac{1}{\omega_1\pm\omega_2}X^2\pm2\omega_1\Omega\Delta z^2\mp 2\omega_1Xz\right\}
\end{equation}
Three of these were non-normalizable and only one was the well behaved normalizable state peaked at $x,z=0$:
The normalizable state is
\begin{equation}\label{normalizable}
\Psi=\exp-\left\{\frac{1}{\Delta}X^2+2\omega_1\Omega\Delta z^2-2\omega_1Xz\right\}
\end{equation}
However if we seek {\it all}  Gaussian eigenstates of the truncated Hamiltonian eq.(\ref{hamxz1}), we find only two states
\begin{equation}
\exp\pm\left\{-\frac{1}{2\Omega}(X-2\Omega^2z)^2 +2\Omega^3z^2\right\}
\end{equation}
Evidently none of these two states is normalizable.
These two Gaussian states  are indeed obtained in the limit $\Delta\rightarrow 0$ from two of the states eq.(\ref{states}).
Thus we seem to find no normalizable Gaussian eigenstates of a quadratic Hamiltonian eq.(\ref{hamxz1}), even though for any finite $\Delta$ a normalizable Gaussian eigenstate exists. This means that the Hamiltonian eq.(\ref{hamxz1}) is not diagonalizable, which indeed can be formally proven \cite{philip_aharon},\cite{pt}.

This conclusion is however a little hasty, as it is based on neglecting the last two terms in eq.(\ref{hamxz}). However, even though these terms are multiplied by $\Delta$, in order to be able to neglect them, we need to be sure that they have vanishing matrix elements in the limit $\Delta\rightarrow 0$. It is easy to see that this is not the case here. Indeed, in the normalizable state eq.(\ref{normalizable}) we have
\begin{equation}
\langle z^2\rangle\sim\langle \pi_X^2\rangle\sim \frac{1}{\Delta}
\end{equation}
so that in fact the last two terms in eq.(\ref{hamxz}) are finite in the limit $\Delta\rightarrow 0$ and therefore cannot be simply discarded.

The normalizable state eq.(\ref{normalizable}) does not disappear without a trace in the degenerate limit, but rather 
 tends to a delta function of $X$
\begin{equation}
\Psi^2(X)\rightarrow\delta(X)
\end{equation}
The action of the Hamiltonian eq.(\ref{hamxz1}) on this state is ambiguous due to the first term in the Hamiltonian. One does obtain this state unambiguously, however as the equal frequency limit of eq.(\ref{normalizable})

Thus on the normalizable states the auxiliary variable $X$ is frozen at zero, while the original variable $z$ fluctuates freely with infinite amplitude. 

Interestingly, this suggests that in a sense the oscillator looses half of its degrees of freedom and also becomes  ``classical". Recall that the variable $X$ is essentially the classical equation of motion for half the original modes of $z$, since $X=\frac{d^2}{dt^2}z+\omega^2 z$. In the limit $\Delta\rightarrow 0$, this quantity is fixed at zero without fluctuations. On the other hand the coordinate $z$ itself fluctuates without restriction. Thus essentially the quantum system becomes a classical oscillator which can oscillate with arbitrary amplitude.

\subsection{Dynamical conformal symmetry}
As an interesting aside, we note that at $\Delta=0$ the theory dynamically develops a conformal symmetry, which is spontaneously broken by normalizable states. For the purpose of this discussion, it is convenient to revert to normalization in which the Hamiltonian is simplest in the limit $\Delta\rightarrow 0$, eq.(\ref{hamxy}).
In the equal frequency limit the Hamiltonian eq.(\ref{hamxy}) is invariant under the following transformation
\begin{equation}\label{conformal}
x\rightarrow x\cosh t  +y\sinh t ; \ \ \ y\rightarrow y\cosh t +x\sinh t ; \ \ \ \ \ z\rightarrow e^{-t}z
\end{equation}
It is natural to refer to this symmetry as conformal.
This symmetry is not obviously present in the Lagrangian eq.(\ref{lagr}). In fact the Lagrangian is multiplied by a constant under the transformation eq.(\ref{conformal}). However, as we have seen in the previous subsection, in the equal frequency limit the dynamics of $z$ is such that on normalizable states it is pinned to satisfy $X=\frac{d^2}{dt^2}z+\omega^2 z=0$. As a result the Lagrangian vanishes for all physically interesting configurations. Scaling of the Lagrangian by a finite factor therefore is indeed a ``dynamical" symmetry in this limit. 

Interestingly this symmetry is spontaneously broken, in the sense that the normalizable ``vacuum", or in fact any of the normalizable physical states, is not invariant under it. 
The wave function of the ``lowest energy", the non-normalizable eigenstate of the operator $a^\dagger$ is indeed invariant under the conformal transformation:
\begin{equation}
\exp\left\{-\frac{1}{2\Omega}\left[x^2-y^2\right]\right\}
\end{equation}
However for the normalizable Gaussian
\begin{equation}
\exp \left\{-\frac{1}{2\Omega}\left[x^2+y^2\right]\right\}\rightarrow \exp\left\{ -\frac{1}{2\Omega}\left[\cosh (2t)\left[x^2+y^2\right]+2\sinh (2t) xy\right]\right\}
\end{equation}
It is clear that any state whose wave function is localized at finite values of $x$ and $y$ is necessarily not invariant under the transformation eq.(\ref{conformal}).
Thus the conformal symmetry is ``spontaneously broken" on normalizable states. Since the representations of conformal group eq.(\ref{conformal}) are infinitely dimensional, the finite energy spectrum is infinitely degenerate. This is of course well known and obvious since adding any number of excitations of the $x$ oscillator and the same number of excitations of the $y$ oscillator does not change the energy in the degenerate limit\cite{philip_aharon,smilga}. It is nevertheless amusing, that this degeneracy can be understood as a spontaneous breaking of conformal symmetry.

\section{Dynamics: Classical vs Quantum}
The dynamics of the classical Pais-Uhlenbeck oscillator is identical to that of two decoupled harmonic oscillators. The variables $x$ and $y$ satisfy the harmonic oscillator equations of motion, and the fact that the energy of the $x$-oscillator is negative is irrelevant, since the energies of each oscillator are separately conserved. 

Quantum mechanically, however the situation is very different. Here the overall sign of energy is reflected in the sign of the phase of the wave function. For the evolution of states which are initially product wave functions $\Psi_1(x)\Psi_2(y)$ this is again unimportant, however it affects strongly the time evolution of ``entangled" states.  The simplest calculation where the quantum mechanical importance of the sign flip for the $x$-oscillator manifests itself, is the propagator of the $z$. It is of course well known, that the UV behavior of the propagator in four derivative theories is much softer than in theories with ordinary kinetic term. The Pais-Uhlenbeck oscillator is the simplest example of this kind.
Although this is a trivial calculation, we present it here for completeness.

\subsection{The propagator}
To calculate the propagator of $z$ we need to calculate the propagator of $x$ and $y$ separately. For $y$ this is the usual harmonic oscillator calculation.

\subsubsection{The $y$ propagator}
The Hamiltonian for the $y$ mode is
\be H = \frac{1}{2}p^2 + \frac{1}{2}\omega_1 y^2\ee
The annihilation operator $a$
\be a = \sqrt{\frac{\omega_1}{2}}(y + \frac{ip}{\omega_1})\ee
evolves in time according to
\be a(t) = a(0) e^{-i \omega_1 t}\ee
For the Feynman propagator:
\be G_y(t)=<T \{ y(t) y(0)\} > = \frac{1}{2\omega_1}<\Theta(t)[a(t)a^\dagger(0) + a^\dagger(t) a(0)] + \Theta(-t)[a^\dagger(0)a(t) + a(0)a^\dagger(t)]>\ee
we have
 \be G_y(t) = \frac{1}{2\omega_1} [\Theta(t) e^{-i \omega_1 t} + \Theta(-t) e^{i \omega_1 t}]\ee 
To perform the Fourier transform, as usual we introduce the regulator which makes the integral converent for large times
\be\label{propy} 
G_y(p)=\frac{1}{2\omega_1} \int dte^{ipt}[\Theta(t) e^{-i \omega_1 t}e^{-\epsilon t} + \Theta(-t) e^{i \omega_1 t}e^{\epsilon t}]=
\frac{i}{p^2-\omega_1^2+ i \epsilon}\ee
This is the standard result, which upon integration over the frequency $p$ gives the equal time expectation value in the vacuum
\be \langle y^2\rangle=\int\frac{dp}{2\pi}G_y(p)=\frac{1}{2\omega_1}\ee

\subsubsection{The $x$ propagator}
The propagator of  $x$ is equally easy to calculate in the physically relevant ``vacuum"- the highest energy state. The Hamiltonian now is
\be H = -\frac{1}{2}p^2 - \frac{1}{2}\omega_2 x^2\ee
and
\be \label{annihil}a = \sqrt{\frac{\omega_2}{2}}(x + \frac{ip}{\omega_2});\ \ \ \ \ \ \ \ \ a(t) = a(0) e^{i \omega_2 t}\ee
The same calculation as before now gives
\be G_x(t)\equiv\langle 0\vert T \{ x(t) x(0)\}\vert 0\rangle=\frac{1}{2\omega_2} [\Theta(t) e^{i \omega_2 t} + \Theta(-t) e^{-i \omega_2 t}]\ee
and
\be G_x(p)=\frac{-i}{p^2-\omega_2^2-i \epsilon}\ee
This differs from eq.(\ref{propy}) by the overall sign and also by the sign of the regulator $\epsilon$. As is easily seen, these two sign changes cancel each other in the calculation of equal time quantities. For example
\be \langle 0\vert x^2\vert 0\rangle=\int \frac{dp}{2\pi}G_x(p)=\frac{1}{2\omega_2} \ee
which is the correct result for the normalizable Gaussian eigenstate of the $x$ oscillator.
\subsubsection{The $z$ propagator}
Finally combining the results for $x$ and $y$, and noting that due to the symmetries of the system the mixed propagator vanishes $\langle x(t)y(0)\rangle=0$, we obtain
\be G_z(p)=\frac{1}{4\Omega\Delta}[G_y(p)+G_x(p)]=\frac{i}{2(p^2-\omega_1^2)(p^2-\omega_2^2)+i\epsilon}\ee
Again, this is the standard result, showing a softened UV behavior, since the propagator of $z$ vanishes much faster for high frequencies than that of a harmonic oscillator. 
This indicates of course, that the time evolution of $z$ is very smooth and has a very small high frequency component. 
\subsubsection{The ``propagator" in the unbounded state}
What happens if we try to calculate the propagator of the $x$ oscillator in the unbounded Gaussian state? Of course, as explained above this calculation is purely formal, since the integrals over this wave function are divergent. Still, formally proceeding as before we can define
\be G^-_x(t)= \langle \Phi\vert T \{ x(t) x(0)\}\vert \Phi\rangle\ee
We still use eq.(\ref{annihil}), but this time it is $a^\dagger$ that annihilates the state $\Phi$. We then formally obtain:
\be G^-_x(t)= -\frac{1}{2\omega_2} [\Theta(t) e^{-i \omega_2 t} + \Theta(-t) e^{i \omega_2 t}]\ee
and 
\be G^-_x(p) = \frac{-i}{p^2-\omega^2+ i \epsilon}\ee
The sign of the regulator $\epsilon$ is now the same as for the positive energy harmonic oscillator, which is simply the reflection of the fact that the state $\vert \Phi\rangle$ is formally the lowest energy state of the system. However this propagator leads to the same paradox of negative norm states as discussed in the previous section. Calculating the equal time expectation value, which should be by definition positive, we find
\be \langle \Phi\vert x^2\vert \Phi\rangle=-\frac{1}{2\omega_2}\ee
This again underscores the point, that non-normalizable states, if manipulated formally,  can be mistaken to have negative norm.

\subsection{Time evolution: the wave function}
It is instructive to see explicitly how the wave function of the system evolves in time. In particular we would like to see the origin of the smooth UV behavior of the Pais-Uhlenbeck system in terms of the time evolution of wave functions. 

We  are mostly interested in the degenerate case $\Delta\rightarrow 0$, and will therefore study time evolution generated by the Hamiltonian
\be H = -\frac{1}{2} \frac{\partial^2}{\partial y^2} + \frac{1}{2} \frac{\partial^2}{\partial x^2} + \frac{1}{2}\Omega^2 y^2 - \frac{1}{2}\Omega^2 x^2. \ee
We want to follow the time dependence of simple quantum averages, like $\langle z^2(t)\rangle$ and $\langle (X(t)+Y(t))^2\rangle$. The first observable is the obvious choice, since it is the fluctuation of the coordinate of the original oscillator, while the second one is the fluctuation of the second order equation of motion. We will choose an initial state such that both these operators have sensible (finite) averages. 

We are not interested in states which are simple product states of  the form $\psi_1(x)\psi_2(y)$. As far as the expectation values of all Hermitian operators go, the evolution of such a product state is identical to that of a state $\psi_1(x)\psi^*_2(y)$ evolved with the positive energy harmonic oscillator.  We will thus be interested in states which are not trivial product states in the variables $x$ and $y$. A simple initial wave function that satisfies these requirements is
\begin{eqnarray}\psi (0)& =& N\exp\left\{-\frac{1}{2}\left[\frac{\Delta\Omega}{\xi^2}(x+y)^2 + \frac{1}{4\Omega\tau^2\Delta}(x-y)^2\right]\right\}\\
&=& N\exp\left\{-\frac{1}{2}\left[(\frac{\Delta\Omega}{\xi^2} + \frac{1}{4\Omega\tau^2\Delta})x^2+(\frac{\Delta\Omega}{\xi^2} + \frac{1}{4\Omega\tau^2\Delta})y^2+2(\frac{\Delta\Omega}{\xi^2} - \frac{1}{4\Omega\tau^2\Delta})xy\right]\right\}.\nonumber \label{wf} \end{eqnarray}
Note that we have scaled out the dependence on the frequency difference $\Delta$ explicitly. Strictly speaking for nonvanishing $\Delta$ we also have to keep the frequencies of the two oscillators in the Hamiltonian different. However the Hamiltonian itself is smooth in the degenerate limit, and it is only the relation between $x,y$ and $z$ that involves divergent coefficients. Thus with the appropriate choice of the wave function we can make $z$ finite also at $\Delta\rightarrow 0$. Specifically, for the state eq.(\ref{wf}) we have
\be \langle z^2\rangle=\tau^2; \ \ \ \ \ \langle (X+Y)^2\rangle=\xi^2\ee
Since the evolution is free, a Gaussian wave function preserves its Gaussian shape at any later time. Thus at any time $t$ we have
\be \psi (t) = N(t)\exp\left[-\frac{1}{2}A(t)x^2-\frac{1}{2}B(t)y^2-C(t)xy\right]. \label{wft}\ee
Acting on this wave function with the Hamiltonian we obtain the evolution of the coefficients
%

\be\label{eqs}\dot{A} = i [A^2 -C^2 -\Omega^2];\ \ \ \dot{B} = i [C^2 -B^2 -\Omega^2]; \ \ \ \dot{C} = i C [A - B].\ee

After some algebra this leads to


\be \dot{C} =  C \frac{\dot{A} + \dot{B}}{A+B}\ee
which is solved by
\be C(t) = \alpha [A(t) + B(t)]\ee
with 
\be \alpha = \frac{C(0)}{A(0) + B(0)} \label{alfa}\ee
Using this result for $C(t)$ in eq.(\ref{eqs}), and defining  $A(t) + B(t) \equiv u(t),\mbox{ }A(t) - B(t) = v(t).$ we have:
\be \dot{u} = i u v \label{udot}\ee
\be \dot{v} = i [(\frac{1}{2} - 2 \alpha^2)u^2 + \frac{1}{2} v^2 - 2\Omega^2] \label{vdot}\ee
with the initial conditions:
\be \label{initcond}u(0) = 2 \left(\frac{\Delta\Omega}{\xi^2} + \frac{1}{4\Omega\tau^2\Delta}\right), \mbox{ } v(0)= 0\ee
It is easy to see that the solution has the form
\be u(t) = \frac{1}{f_+ + f_- \cos{2\Omega t}}, \ \ \ \ \  v(t) = \frac{- i 2f_- \Omega \sin {2\Omega t}}{f_+ + f_- \cos{2\Omega t}}\label{ansatz}
\ee
where $f_\pm$ are constants determined by the equations of motion and the initial conditions. After some algebra, for the initial conditions eq.(\ref{initcond}) we obtain
\be f_\pm=\frac{\Delta}{\Omega}(\pm 1+\Omega^2\xi^2\tau^2)\frac{1}{\xi^2+4\Omega^2\Delta^2\tau^2}    \ee
and
\begin{eqnarray} A(t)&=& \frac{\Omega}{2\Delta}\frac{(\xi^2+4\Omega^2\Delta^2\tau^2)-i2\Delta(1-\Omega^2\xi^2\tau^2)\sin 2\Omega t}{(1-\cos 2\Omega t)+\Omega^2\xi^2\tau^2(1+\cos 2\Omega t)} \\
B(t)&=& \frac{\Omega}{2\Delta}\frac{(\xi^2+4\Omega^2\Delta^2\tau^2)+i2\Delta(1-\Omega^2\xi^2\tau^2)\sin 2\Omega t}{(1-\cos 2\Omega t)+\Omega^2\xi^2\tau^2(1+\cos 2\Omega t)} \nonumber\\
C(t)&=& -\frac{\Omega}{2\Delta}\frac{\xi^2-4\Omega^2\Delta^2\tau^2}{(1-\cos 2\Omega t)+\Omega^2\xi^2\tau^2(1+\cos 2\Omega t)} \nonumber\end{eqnarray}

The time dependent probability density can be written as:
\be \psi^\dagger \psi = N^2 \exp\left[ 
 - \frac{\Omega}{2\Delta}(x+y)^2\frac{4\Omega^2\Delta^2\tau^2}{(1-\cos 2\Omega t)+\Omega^2\xi^2\tau^2(1+\cos 2\Omega t)}- \frac{\Omega}{2\Delta}(x - y)^2  \frac{\xi^2}{(1-\cos 2\Omega t)+\Omega^2\xi^2\tau^2(1+\cos 2\Omega t)}\right] \ee
 Thus we find
\begin{eqnarray} \label{timeav}<z^2(t)>& =& \frac{1}{2}\left[\frac{1}{\Omega^2\xi^2}(1-\cos 2\Omega t)+\tau^2(1+\cos 2\Omega t)\right]\\
\langle(X(t)+Y(t))^2\rangle&=&\frac{1}{2}\left[\frac{1}{\Omega^2\tau^2}(1-\cos 2\Omega t)+\xi^2(1+\cos 2\Omega t)\right]\nonumber\end{eqnarray}
These expressions are  notable for their absence of features. Normally one expects that if the initial state is very far from the vacuum, the evolution should delocalize it in a short time, so that the amplitude of the fluctuation of the coordinates should become very large. This is exactly what happens in the standard positive Hamiltonian harmonic oscillator, as we will demonstrate in the next subsection. However eq.(\ref{timeav}) shows that in the Pais-Uhlenbeck system both interesting averages evolve smoothly in time on the scale determined by the initial state averages. Clearly, if both $\tau$ and $\xi$ are finite, the averages stay finite throughout the evolution. This is despite the fact, that the ``vacuum" of the system is such that $\tau^2\propto 1/\Delta\rightarrow\infty$, $\xi^2\propto\Delta\rightarrow 0$, as discussed in the previous section. If we start the system ``close" to its vacuum state, that is with $\xi^2\propto 1/\tau^2\propto\Delta$, it is still true that at all times parametrically the averages are the same, fluctuation with the amplitude proportional to the initial average. Thus it does not matter, if the system starts off far from the vacuum, or close to it, the evolution is smooth and the averages at all times are proportional to those in the initial state.

To underscore that this is very different from the standard harmonic oscillator, we perform the same exercise as above for the two decoupled oscillator systems.

\subsection{The baseline: oscillators with positive energy}

We now consider time evolution generated by
\be  H = -\frac{1}{2} \frac{\partial^2}{\partial y^2} - \frac{1}{2} \frac{\partial^2}{\partial x^2} + \frac{1}{2}\Omega^2 y^2 + \frac{1}{2}\Omega^2 x^2 \ee
For a Gaussian wave function eq.(\ref{wft}) the evolution of the parameters $A(t)$, $B(t)$ and $C(t)$ is given by:

\be \dot{A} =  i [-C^2 - A^2 + \Omega^2] \label{aa};\ \ \ \dot{B} =  i [-C^2 - B^2 + \Omega^2]\label{bb};\ \ \  \dot{C} = -i C(A+B) \label{cc}\ee

This is simplified for our initial state where $A(t)$ and $B(t)$ stay equal for all times:
\be \dot{A} =  i [-C^2 - A^2 + \Omega^2] ; \ \ \ \ 
\dot{C} = -2iA C, \label{ac}\ee
with initial conditions given as in \ref{wf}.
This is solved by 
\be C= \frac{1}{f + g \cos{(2\Omega t + \phi)}}; \ \ \ \ 
A= \frac{i g \Omega \sin{(2\Omega t + \phi)}}{f + g \cos{(2\Omega t + \phi)}} \ee 
provided
\be f^2 - g^2 = \frac{1}{\Omega^2} \ee

Imposing the initial conditions,
\be A(0) = \frac{\Delta\Omega}{\xi^2} + \frac{1}{4\Omega\tau^2\Delta} = \frac{i g \Omega \sin{\phi}}{2(f + g \cos {\phi})}; \ \ \ \ \ 
 C(0) = \frac{\Delta\Omega}{\xi^2} - \frac{1}{4\Omega\tau^2\Delta} = \frac{1}{f + g \cos{\phi}}.\ee
we find
\be f= \frac{2\Delta}{\Omega}\frac{\Omega^2\tau^2\xi^2-1}{4 \Omega^2 \Delta^2\tau^2-\xi^2};\ \ \ 
g \sinh\Phi=-\frac{1}{\Omega} \frac{4\Omega^2\Delta^2\tau^2 +\xi^2}{4\Omega^2\Delta^2\tau^2 -\xi^2}; \ \ \ 
 g \cosh\Phi=2\Delta\frac{\Omega^2\tau^2\xi^2+1}{4\Omega^2\Delta^2\tau^2 -\xi^2}\ee

where $\Phi=i\phi$.
Finally, the solution for our initial conditions is
\begin{eqnarray} A(t) = B(t) &=& \frac{\Omega}{2\Delta}\frac{(4\Omega^2\Delta^2\tau^2 +\xi^2)\cos2\Omega t+i2\Delta(\Omega^2\tau^2\xi^2+1)\sin 2\Omega t}{\Omega^2\tau^2\xi^2(1+\cos 2\Omega t)-(1-\cos 2\Omega t)+i(2\Omega^2\Delta\tau^2 +\frac{\xi^2}{2\Delta})\sin 2\Omega t}\nonumber\\
C(t) &=&\frac{\Omega}{2\Delta}\frac{4\Omega^2\Delta^2\tau^2 -\xi^2}{\Omega^2\tau^2\xi^2(1+\cos 2\Omega t)-(1-\cos 2\Omega t)+i(2\Omega^2\Delta\tau^2 +\frac{\xi^2}{2\Delta})\sin 2\Omega t}
\end{eqnarray}
For small $\Delta$ we expand these expressions to second nontrivial order
\begin{eqnarray}
A(t)=B(t)&=&i\Omega\tan 2\Omega t+\frac{2\Omega\Delta}{\xi^2}\frac{(\Omega^2\tau^2\xi^2-1)\cos 2\Omega t+(\Omega^2\tau^2\xi^2+1)\cos 4\Omega t}{\sin^2 2\Omega t} \nonumber\\
C(t)&=&-i\frac{\Omega}{\sin 2\Omega t}-\frac{2\Omega\Delta}{\xi^2}\frac{\Omega^2\tau^2\xi^2(1+\cos 2\Omega t)-(1-\cos 2\Omega t)}{\sin^2 2\Omega t}
\end{eqnarray}
Generically at arbitrary time we have
\be Re[A+C]\propto Re[A-C]\propto\frac{\Delta}{\sin^22\Omega t}\ee
and thus 
\be\langle (x-y)^2\rangle\propto \langle (x+y)^2\rangle \propto\frac{\sin^22\Omega t}{\Delta}\label{averages}
\ee
This is precisely what one normally expects. Our initial state is very far away from the ground state. It was chosen in such a way that the center of mass coordinate $x+y$ had large fluctuations, $O(1/\Delta)$, whereas the relative coordinate $x-y$ had small fluctuations $O(\Delta)$. One expects a state like this to expand very quickly and become delocalized in all coordinates. Indeed eq.(\ref{averages}) displays precisely this feature: the relative coordinate fluctuates with amplitude of order $1/\Delta$ almost all the time, except for a very short time interval 
 $\delta t\propto \Delta$ within every period of evolution. 

Thus indeed, we see that the time evolution of the Pais - Uhlenbeck oscillator is smoother than that of a system of decoupled harmonic oscillators, in the sense that the averages in the Pais-Uhlenbeck case fluctuate on the scale given by the initial state and do not develop additional large variations throughout the evolution.

\section{A simple unitary interaction}
We have seen that the quantum evolution of the Pais-Uhlenbeck oscillator is unitary. This is not very surprising, nor very exciting since the two second order degrees of freedom in this case are decoupled, and each one follows a Harmonic oscillator evolution. In fact the system has two conserved quantum numbers - not just the total energy, but also the energy of each individual oscillator is conserved. For this reason the classical motion in the $X,Y$ plane is bounded and the quantum evolution is unitary.

A more interesting and general question is whether interacting systems with ghosts can be unitary.
The worry is clear. We have a Hamiltonian which is unbounded neither from above nor from below, and once the two modes $x$ and $y$ are allowed to interact, there is a real and present danger that the system can develop an instability, where both $x$ and $y$ run away to infinity even though the total energy stays conserved. 

In the quantum mechanical context one can pose the following question: does  a system of coupled ``particle" and ``ghost" degrees of freedom possess normalizable eigenstates. If the answer is affirmative, such system enjoys unitary quantum evolution, since the probability to find the system in finite volume does not decrease with time \footnote{One should qualify this statement slightly. An initial state that has a finite but nonunit projection onto a subspace spanned by normalizable eigenstates will leak probability initially. This leakage will stop after a while and the rest of the evolution will be unitary, preserving the part of the total probability associated with the normalizable subspace. Such behavior is physically perfectly admissible and we will refer to it as unitary disregarding any initial transient leakage of probability}. If this is not the  case, such systems would not allow for unitary quantum mechanical evolution and probability would leak out completely through the boundaries in a finite amount of time.

The aim of this section is to present a simple example of a model, which remains unitary even though it contains interacting particle and ghost degrees of freedom \footnote{ We note that an example of a stable supersymmetric system with ghosts was discussed in \cite{smilga1}.}. 
Let us add to our Hamiltonian a quartic interaction of the form
\begin{equation}\label{coupled}
H=\frac{1}{2}\pi_y^2+\frac{1}{2}\omega_1^2y^2+\lambda_1y^4-\frac{1}{2}\pi_x^2-\frac{1}{2}\omega_2^2x^2-\lambda_2x^4+\mu x^2y^2
\end{equation}
For definiteness we choose $\mu>0$. At $\mu=0$ the theory is clearly unitary, as the particle and ghost degrees of freedom are decoupled, and evolution of each one separately is unitary in exactly the same sense as for the Pais - Uhlenbeck oscillator.

The question about stability can be asked already on the classical level. It was noted in \cite{trodden} and also \cite{smilga}, that some systems of this kind allow for classically stable solutions, namely oscillatory solutions for which the amplitude does not grow without bound as a function of time. Specifically ref.\cite{trodden} studied numerically the evolution of eq.(\ref{coupled}) for  $\lambda_{1,2}=0$ and found that the classical behavior of the system is stable as long as the initial energy stored in the oscillators is not too large. Denoting the initial displacement of the oscillators from the equilibrium by $M$, ref.\cite{trodden} found that for $M^2<M^2_c=\frac{1}{\mu}\omega_2^2$ the behavior is oscillatory, while for $M^2>M_c^2$ the amplitude of oscillations grows without bound.
The addition of the quartic self interaction $\lambda_{1,2}$ further stabilizes the system. We have repeated the numerical exercise of \cite{trodden} for the system eq.(\ref{coupled}), and have found a similar behavior in a wider range of parameters. In fact as long as the coupling $\mu$ remains small $\mu\ll\lambda_{1,2}$ we did not see classical instability  for any initial conditions that we have tried. Examples of evolution for several initial conditions are given in Fig.1. This suggests that when the interaction is weak enough, the classical system is absolutely stable, although it is not possible to prove such a statement by numerical methods.

\begin{figure}[ht!]
     \begin{center}
        \subfigure[$x_0=21, y_0=20, \dot{x}_0=25, \dot{y}_0=29$]{%
            \label{fig:first}
            \includegraphics[width=0.4\textwidth]{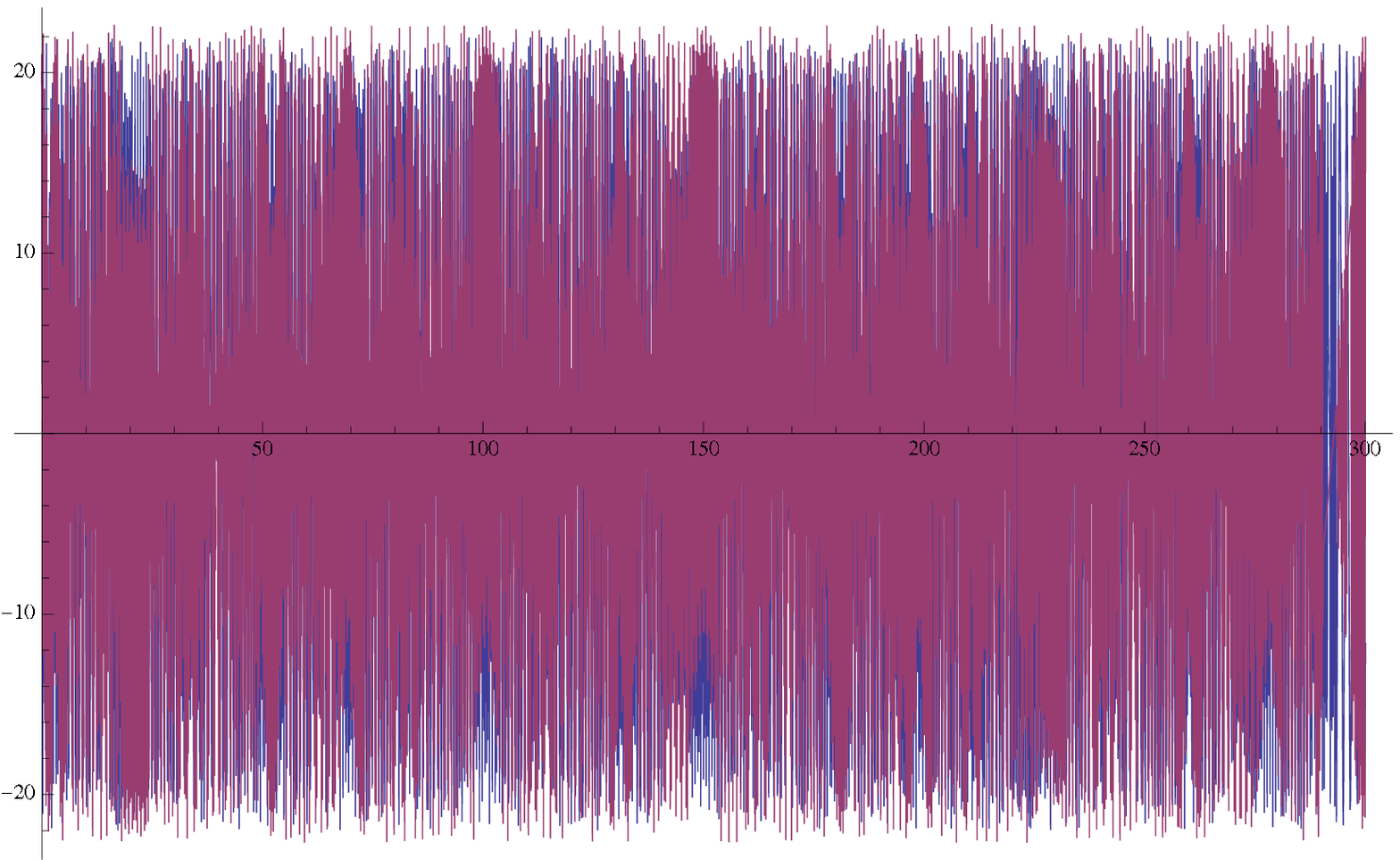}
        }%
        \subfigure[$x_0=21, y_0=20, \dot{x}_0=25, \dot{y}_0=29$]{%
           \label{fig:second}
           \includegraphics[width=0.4\textwidth]{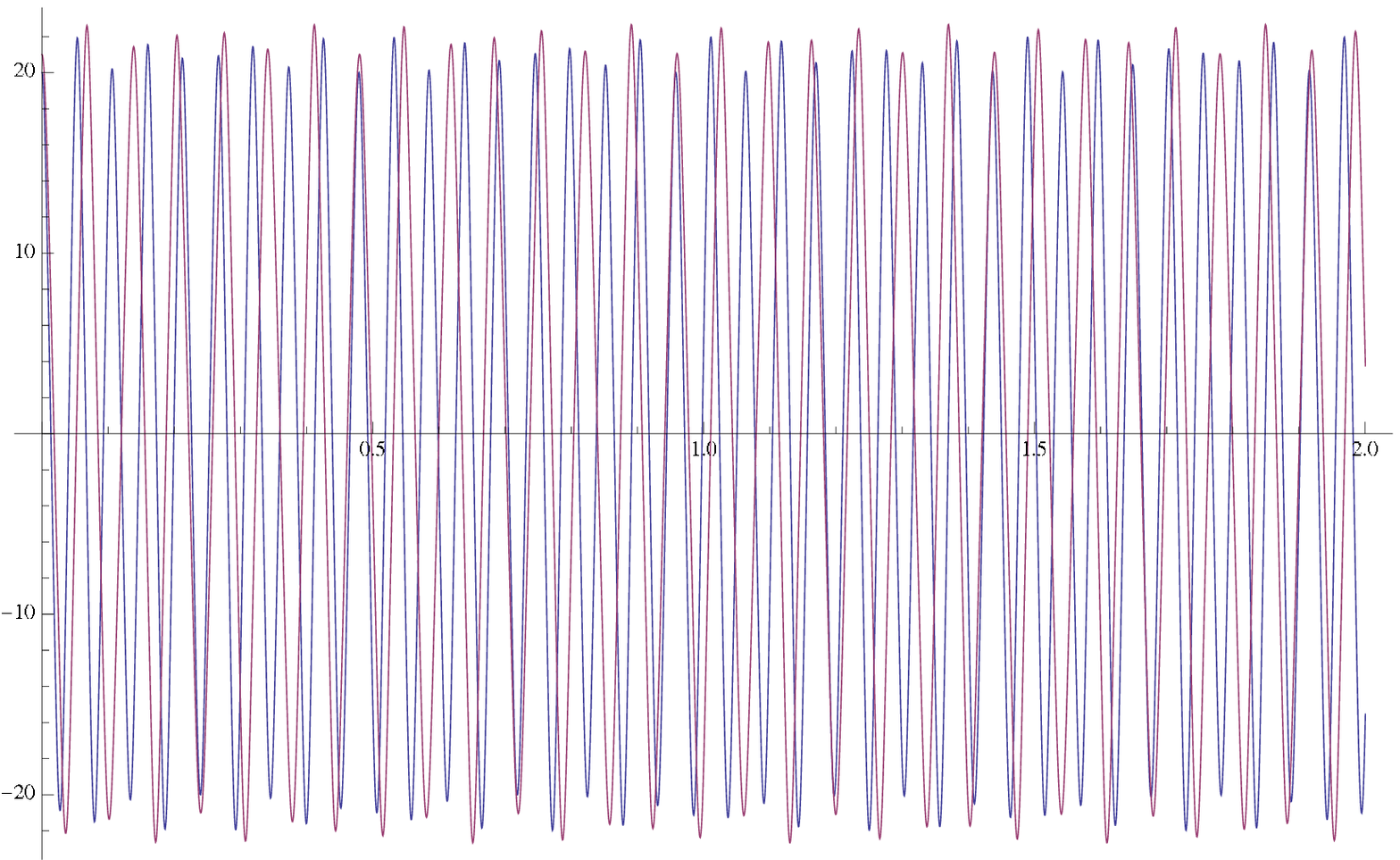}
        }\\ 
        \subfigure[$x_0=1, y_0=1, \dot{x}_0=0, \dot{y}_0=0$]{%
            \label{fig:third}
            \includegraphics[width=0.4\textwidth]{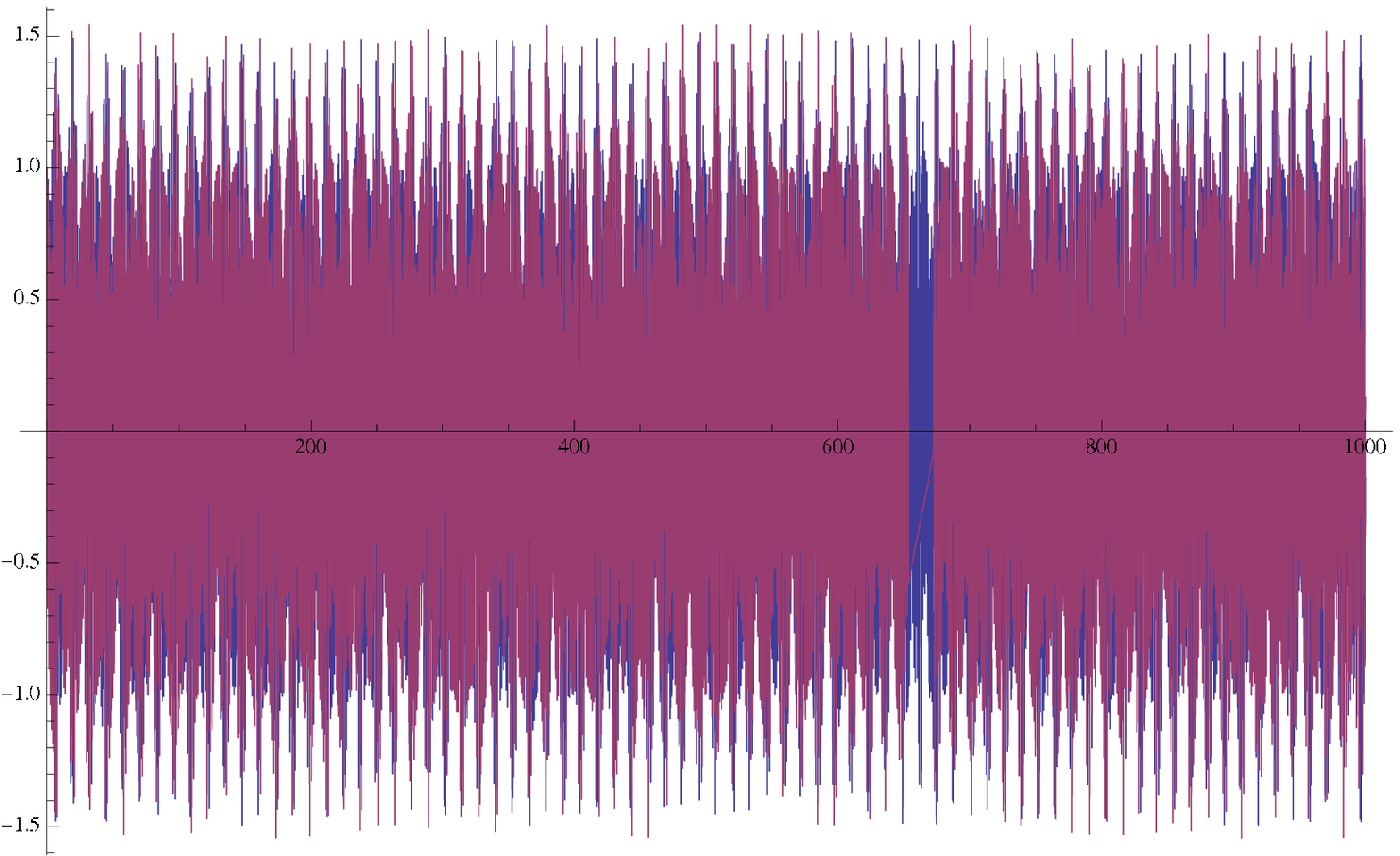}
        }%
        \subfigure[$x_0=1, y_0=1, \dot{x}_0=0, \dot{y}_0=0$]{%
            \label{fig:fourth}
            \includegraphics[width=0.4\textwidth]{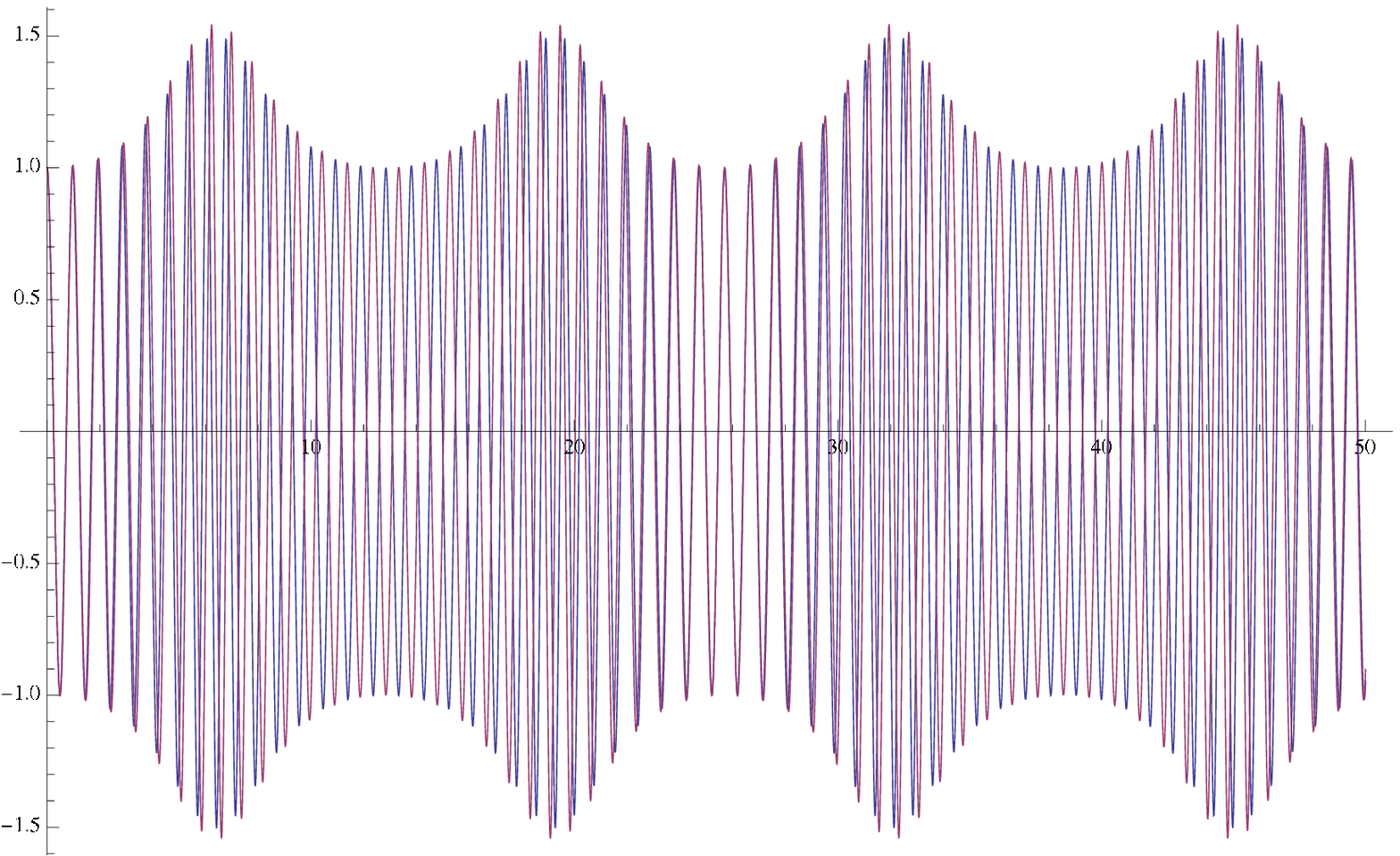}
        }%

    \end{center}
    \caption{%
     Typical time evolution of $x$ (red) and $y$ (blue) for different initial conditions. The parameters are chosen as $\omega_1 = 3, \lambda_1=10, \omega_2=5, \lambda_2=7, \mu=3$. The evolution is plotted over two time intervals to show the detailed structure of time dependence and to demonstrate the absense of instability over very long times.       }%
          \label{fig:subfigures}
\end{figure}

Note, that in order for the quantum system to be unitary, its classical counterpart has to have stable evolution for arbitrary initial conditions. Otherwise quantum tunneling will connect stable and unstable regions of the phase space and will inevitably lead to violation of unitarity. This is the situation, for example in the upside down Mexican hat potential $U(x)=-\lambda (x^2-x^2_0)^2$. Classical solutions with total energy $-\lambda x_0^4<E<0$ and initial displacement $|x|<x_0$ are regular. However quantum mechanically the system is non-unitary due to finite probability of tunneling into the unbounded region $|x|>x_0$.

In the present case, one can give an argument that the theory remains stable, at least in a limited range of parameters. Let us consider the limit $\omega_1\ll\omega_2$. In this case one can use the  classical Born-Oppenheimer approximation. Since $y$ oscillates much faster than $x$, one can consider the motion of $y$ in the background of fixed $x$.
Thus for given $x$ the dynamics of $y$ is given simply by an anharmonic oscillator with the frequency, which for large $x$ behaves as $\omega^2=\mu x^2$. This is clearly a well defined bounded motion. The dynamics of $x$ is affected by the average value of $y^2$ for a given trajectory. Given the initial energy $E$ stored in the mode $y$, we have (for large $x$, which is the interesting and potentially dangerous regime) $\bar y^2\propto E/\mu x^2$. The dynamics of $x$ then is governed by the effective potential
\begin{equation}
\frac{1}{2}\omega_2^2x^2+\lambda_2x^4-\mu\bar y^2x^2=\frac{1}{2}\omega_2^2x^2+\lambda_2x^4-E
\end{equation}
Thus the dynamics of $x$ in this approximation is unaffected by $y$ and is bounded and stable. A similar argument can be given for the opposite case $\omega_1\gg\omega_2$. Thus at least when the two frequencies are very different there is no instability for arbitrary initial conditions. In this case one expects that the quantum theory is well defined and unitary in the sense explained above.

In the next subsection we present another line of reasoning supporting the same conclusion for small $\mu$. 

\subsection{Asymptotics of Eigenfunctions for Small $\mu$ }

One way to establish that a quantum theory has normalizable eigenstate is to find asymptotics of eigenfunctions for large values of coordinates $x$ and $y$.

As usual, we introduce an eikonal $S$ via
\be\label{eikonal} \Psi = N e^{-S(x,y)}\ee
If the eikonal is positive and divergent for large values of the coordinates, the wave function is normalizable. 
For large values of $S$, $|x|$ and $|y|$  it satisfies the following ``semiclassical" equation:
\be -\frac{1}{2} \left(\frac{\partial S}{\partial y}\right)^2 + \frac{1}{2}\left(\frac{\partial S}{\partial x}\right)^2 + \lambda_1 y^4 - \lambda_2 x^4 + \mu x^2 y^2 = 0\ee
We will not attempt to solve this equation in full generality, but rather explore the behavior of $S$ for small values of $\mu$. For $\mu=0$ the solution is simply a sum of the solutions for two decoupled degrees of freedom:
\be S_0 (x,y) = \frac{\sqrt{2\lambda_1}}{3} |y|^3 + \frac{\sqrt{2\lambda_2}}{3} |x|^3  \ee
The crucial point is that the structure of the potential is such that for $\mu\ll\lambda_i$, the perturbation is smaller than the leading order potential for generic large values of $x$ and $y$.  This is of course very different from the standard perturbation theory around a harmonic oscillator potential, where  a perturbation is usually bigger than the unperturbed potential for large values of the coordinate. Thus although the standard perturbation theory around a Harmonic potential is asymptotic, we expect the perturbation theory in $\mu$ to have a finite radius of convergence.

Let us therefore solve eq.(\ref{eikonal}) perturbatively.
Let $S = S_0 + S_1$, where $S_1 \propto \mu $. We first solve the equation for $x,y > 0$. To first order in $\mu$ we have:
\be -\sqrt{2\lambda_1} y^2\frac{\partial S_1}{\partial y} + \sqrt{2\lambda_2} x^2 \frac{\partial S_1}{\partial x} + \mu x^2 y^2 =0\ee

Changing variables $ \bar{x} = \frac{1}{\sqrt{2\lambda_2}x}, \mbox{ } \bar{y} = \frac{1}{\sqrt{2\lambda_1}y}$ and defining  $x^{\pm} = \bar{x} \pm \bar{y}$ the equations becomes simple
%
%
%
\be \frac{\partial S_1}{\partial x^-} = \frac{4 \mu}{ \lambda_1 \lambda_2}\frac{1}{(x^{+2} - x^{- 2})^2}\ee

A well behaved solution to this equation is:

\be S_1 (x^+, x^-) = \frac{4 \mu}{\lambda_1 \lambda_2} \frac{1}{2x^{+3}}\left[-\frac{x^+ x^-}{x^{-2} - x^{+2}} + arctanh\left(\frac{x^-}{x^+}\right)\right]\ee

In terms of the original variables, the solution can be written as:

\be S_1(x>0, y>0) = \sqrt{2} \mu x^2 y^2 \frac{\sqrt{\lambda_1} y - \sqrt{\lambda_2} x}{(\sqrt{\lambda_1} y + \sqrt{\lambda_2} x)^2} + 2\sqrt{2} \mu \sqrt{\lambda_1 \lambda_2} \frac{x^3 y^3}{(\sqrt{\lambda_1} y + \sqrt{\lambda_2} x)^3} \log({\sqrt{\frac{\lambda_1}{\lambda_2}}\frac{y}{x}})  \ee
Extending the solution to other regions of the plane we find
\be S_1(x, y) = \sqrt{2} \mu x^2 y^2 \frac{\sqrt{\lambda_1} |y| - \sqrt{\lambda_2} |x|}{(\sqrt{\lambda_1} |y| + \sqrt{\lambda_2} |x|)^2} + 2\sqrt{2} \mu \sqrt{\lambda_1 \lambda_2} \frac{|x|^3 |y|^3}{(\sqrt{\lambda_1} |y| + \sqrt{\lambda_2} |x|)^3} \log({\sqrt{\frac{\lambda_1}{\lambda_2}}\frac{|y|}{|x|}})  \ee

%




As expected, the correction $S_1$ is smaller than $S_0$ at large values of the arguments, and thus the asymptotics of the wave function is determined by $S_0$. Thus we find that for small $\mu$ our model quantum mechanically has normalizable eigenstates, and therefore unitary evolution.

\section*{Acknowledgments}
The work of AK and IBI is supported by DOE grant DE-FG02-92ER40716.


\end{document}